\relax

\def\LNCSmode{1} 
\ifnum\LNCSmode=1
    \documentclass{llncs}
    \usepackage{times,fullpage}
\else
    \documentclass[letterpaper]{article} 
    \usepackage{aaai21} 
    \usepackage{times} 
\fi

\usepackage{helvet} 
\usepackage{courier} 
\usepackage[hyphens]{url} 
\usepackage{hyperref}
\usepackage{graphicx} 
\usepackage{graphicx}  
\usepackage{caption} 
\usepackage{lineno}
\usepackage{booktabs}
\usepackage{multirow}
\usepackage{siunitx}
\usepackage{xcolor}

\title{The LOB Recreation Model: Predicting the Limit Order Book from TAQ History Using an Ordinary Differential Equation Recurrent Neural Network
    }

\ifnum\LNCSmode=1
    \author {
        Zijian Shi
        \and
        Yu Chen
        \and
        John Cartlidge
        \thanks{This is a preprint manuscript accepted for publication in the 35th AAAI Conference on Artificial Intelligence (AAAI-2021).}
        }
        
    \institute{Department of Computer Science, University of Bristol, Bristol, UK.\\
    \email{zijian.shi@bristol.ac.uk, yc14600@bristol.ac.uk, john.cartlidge@bristol.ac.uk}
    }
\else
    \author {
        Zijian Shi, \textsuperscript{\rm 1}
        Yu Chen, \textsuperscript{\rm 1}
        John Cartlidge \textsuperscript{\rm 1} \\}
    
    \affiliations {
        \textsuperscript{\rm 1} Department of Computer Science, University of Bristol\\
        Bristol, BS8 1UB, United Kingdom \\
        zijian.shi@bristol.ac.uk, yc14600@bristol.ac.uk, john.cartlidge@bristol.ac.uk}
\fi

\begin{document}

\maketitle

\begin{abstract}
In an order-driven financial market, the price of a financial asset is discovered through the interaction of orders - requests to buy or sell at a particular price - that are posted to the public limit order book (LOB).
Therefore, LOB data is extremely valuable for modelling market dynamics. However, LOB data is not freely accessible, which poses a challenge to market participants and researchers wishing to exploit this information.
Fortunately, trades and quotes (TAQ) data - orders arriving at the top of the LOB, and trades executing in the market - are more readily available. 
In this paper, we present the LOB recreation model, a first attempt from a deep learning perspective to recreate the top five price levels of the LOB for small-tick stocks using only TAQ data. 
Volumes of orders sitting deep in the LOB are predicted by combining outputs from: (1) a history compiler that uses a Gated Recurrent Unit (GRU) module to selectively compile prediction relevant quote history; (2) a market events simulator, which uses an Ordinary Differential Equation Recurrent Neural Network (ODE-RNN) to simulate the accumulation of net order arrivals; and (3) a weighting scheme to adaptively combine the predictions generated by (1) and (2). 
By the paradigm of transfer learning, the source model trained on one stock can be fine-tuned to enable application to other financial assets of the same class with much lower demand on additional data. Comprehensive experiments conducted on two real world intraday LOB datasets demonstrate that the proposed model can efficiently recreate the LOB with high accuracy using only TAQ data as input.
\end{abstract}

\section{Introduction}
\noindent 
The majority of today's financial markets are order-driven, with traders submitting orders -- requests to buy or sell some quantity at a particular price -- to a public limit order book (LOB). 
Orders resting in the LOB form the instantaneous supply and demand for an asset, with buy orders (or {\em bids}) representing quantity demanded at each price level and sell orders (or {\em asks}) representing quantity supplied at each price level. 
Orders are matched using the continuous double auction (CDA) mechanism such that a buyer or seller can submit an order at any time and a trade execution will occur whenever prices cross; i.e., when an ask (order to sell) price is less than or equal to a bid (order to buy) price. 
The CDA mechanism and LOB determine the market microstructure  \cite{o1997market} and enable a price formation process such that the instantaneous price of a financial asset is entirely determined by the contemporaneous demand and supply in the market. 
In this manner, prices fluctuate freely and evolve over time as orders are submitted, cancelled, and executed in the market.

Given the tight association between the LOB of a financial asset and its price evolution, LOB data is a valuable resource for financial practitioners and academic researchers; enabling back-testing of trading algorithms, offering the potential to forecast price movements, and providing a resource to understand market dynamics.
However, because of this value, full LOB data is often supplied at considerable cost, with subscription fees reaching tens of thousands of dollars annually.\footnote{\url{http://www.nasdaqtrader.com/Trader.aspx?id=DPUSdata}} 
In contrast, trades and quotes (TAQ) data -- trade executions and orders at the top of the LOB, which represent the current best bid and ask (i.e., the highest priced bid, and the lowest priced ask) -- is much more readily available and often freely reported as an asset’s current price and last trade price \cite{bouchaud2018trades}.  This situation provides strong motivation to attempt a recreation of the LOB at depth using only the TAQ data stream as input.

\paragraph{Contribution}

This paper presents a novel \textit{limit order book recreation model} (LOBRM), where order volumes resting at different price levels of the LOB are predicted using only TAQ data. 
The LOBRM consists of three components: (i) a history compiler (HC), which uses a Gated Recurrent Unit (GRU) module \cite{cho2014learning} to compile the evolving history of quote volumes; (ii) a market events simulator (ES), which uses an Ordinary Differential Equation Recurrent Neural Network (ODE-RNN) module with gating control \cite{rubanova2019latent} to encode TAQ history into a continuous latent state that is then decoded by a Multi-Layer Perceptron (MLP) into vectors of order arrival rates at different price levels; and (iii) an adaptive weighting scheme (WS), which fuses the output from (1) and (2) to make a final prediction. 

\section{Background and Related Work}\label{sec:background}

\paragraph{The Limit Order Book (LOB)} 

The continuous double auction (CDA) mechanism used by most major financial markets enables market participants to enter buy and sell orders at any time. 
A {\em limit} order $L(t,s,d,v,p)$ specifies a time of submission $t$, a stock ID $s$, a direction $d$, a volume to trade $v$, and a limit price to trade $p$, which for a buy order (i.e., a {\em bid}) is the highest acceptable trade price and for a sell order (i.e., an {\em ask}) is the lowest acceptable trade price. 
Each limit order $L$ enters the limit order book (LOB) for a given stock $s$. 
The LOB contains a bid list and an ask list, each sorted by price-time priority such that the bid at the front of the bid list (i.e., the {\em best bid}) has the highest price, $p^{b(1)}$, and the ask at the front of the ask list (i.e., the {\em best ask}) has the lowest price, $p^{a(1)}$. 
When a new bid, $b$, is entered with price $p^{b}$, it will execute against the best ask if $p^{b} \geq p^{a(1)}$, else $b$ will enter the LOB in price ordered position. 
Likewise, when a new ask, $a$, is entered with price $p^{a}$, it will execute against the best bid if $p^{a} \leq p^{b(1)}$, else $a$ will enter the LOB in price ordered position. 

\begin{figure}[tb]
\centering
\includegraphics[width=0.85\linewidth]{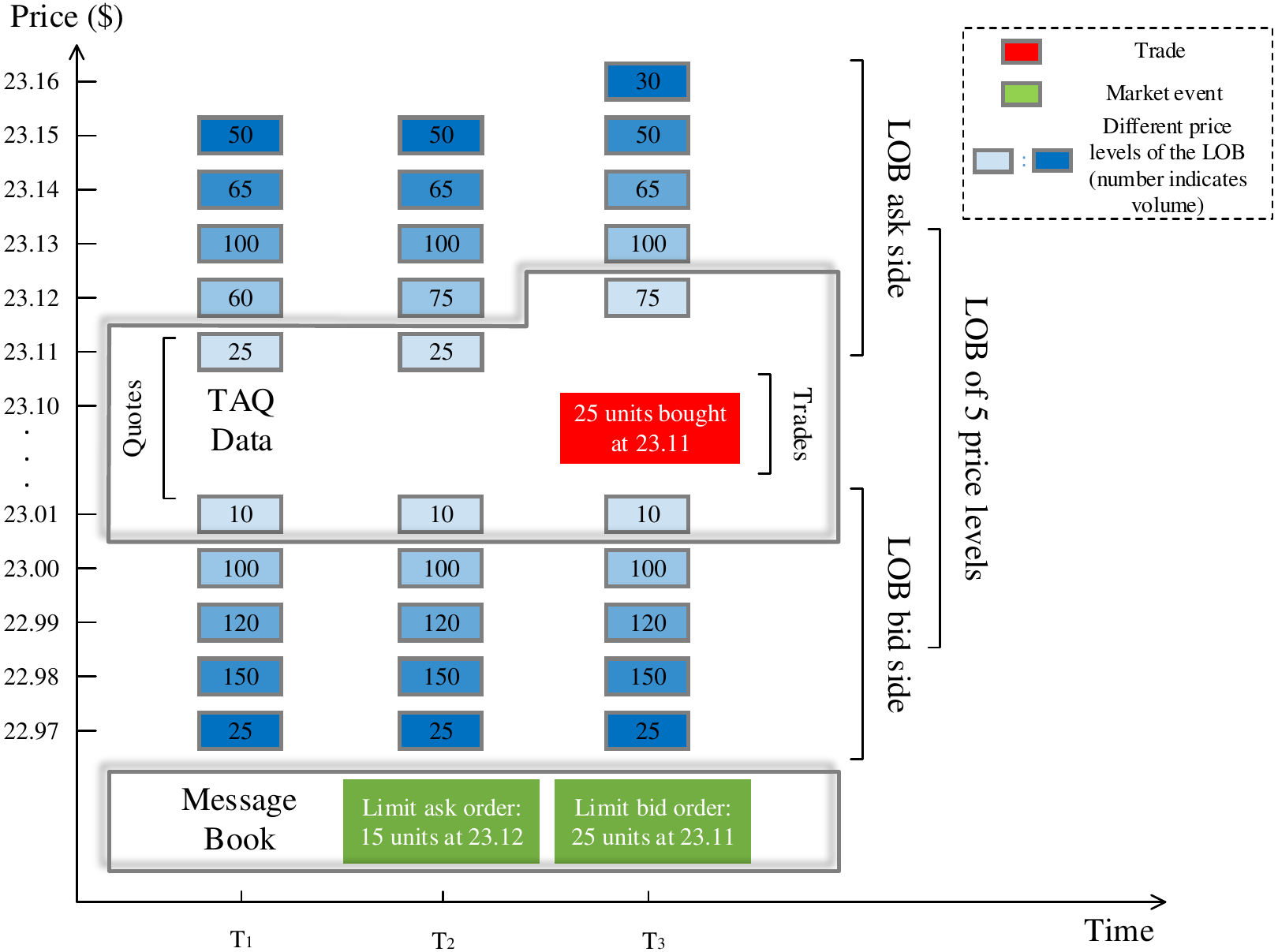} 
\caption{Evolution of a LOB containing 5 price levels, with corresponding message book showing events.}
\label{fig:lob}
\end{figure}

Fig.~\ref{fig:lob} presents a schematic of a LOB with 5 price levels evolving over time. 
Initially, at time $T_1$, the best bid has price \$23.01 with volume 10 and the best ask has price \$23.11 with volume 25. 
The next event (time $T_2$) is a new ask order of price \$23.12 with volume 15. 
This ask price is too high to execute and so the order rests in the LOB on the ask side at the second price level. 
The following event (time $T_3$) is a new bid with price \$23.11 and volume 25. 
This order executes in full against the best ask, which moves the best ask price to \$23.12. 

The full LOB data contains a record of the volume at all price levels in the order book after every event (a trade, or a new order entry or cancellation), along with an affiliated message book that records all market events.
In comparison, trades and quotes (TAQ) data, which is generally more freely available than LOB data, includes only the best bid and ask and a record of trades. 
Although empirical studies have indicated that the first level of the LOB accounts for approximately 80\% of future price movements \cite{cao2009information}, there remains value to exploit in the information stored deeper in the LOB \cite{biais1995empirical,muni2015order}. This provides motivation for modelling the full LOB from TAQ data.

\paragraph{Modelling the LOB}

There have been multiple attempts at stochastic modelling of the LOB. In these works, the evolution of the LOB is described as a higher-order Markov system, with the arrival of events following a particular probabilistic process \cite{cont2013price,blanchet2017unraveling}.
\cite{cont2010stochastic} proposed a continuous-time stochastic model for LOB dynamics, in which the occurrences of events such as arrival and cancellation of limit orders were presumed to follow independent Poisson processes conditioned on the current state of the LOB. 
In the work of \cite{vvedenskaya2011non}, LOB dynamics were formulated as a discrete time Markov process of which the deterministic dynamical system is controlled by nonlinear ODEs. 
Stochastic models are capable of capturing long-term and mid-term empirical features in LOB evolution. However, in the noisier high frequency domain, stochastic models with relatively few parameters are not a good fit as their strong probabilistic assumptions of market events are likely to fail. 

In recent years, there has been an emergence of research using deep learning to model and exploit the LOB. 
One significant study by \cite{sirignano2019universal} on a comprehensive pool of 500 stocks reveals that features learned by a Long Short-Term Memory (LSTM) network are universal for all stocks, and can be used to explain price formation mechanisms. In particular, their universal LSTM model is able to predict next mid-price movement with around 70\% accuracy across all 500 stocks. \cite{sirignano2019universal} also demonstrate that deep learning models suffer less from problems that exist in statistical models of the LOB, such as regime drift. 
Other researches include feature design of the LOB \cite{passalis2019deep}, exploiting LOB dynamics for price trend prediction \cite{zhang2019deeplob}, and generating simulated order flows \cite{li2019generating}. 

However, most prior work assumes that full LOB data is available for model training, but unfortunately this is often not the case. Financial market simulators (e.g., {\em Santa Fe Artificial Stock Market} \cite{arthur1996asset}, {\em Exchange Portal} \cite{stotter2014behavioural}, and {\em Bristol Stock Exchange} \cite{cliff2018open}) offer the opportunity to synthesise LOB data, but this data may not accurately characterise the real world. 
Therefore, the best approach for accurately recreating LOB data is to forecast the LOB using real world TAQ data.
In this paper, we take a deep learning approach for LOB recreation, using an ODE-RNN as the core component. 

\begin{figure}[t]
    \centering
    \includegraphics[width=0.5\linewidth]{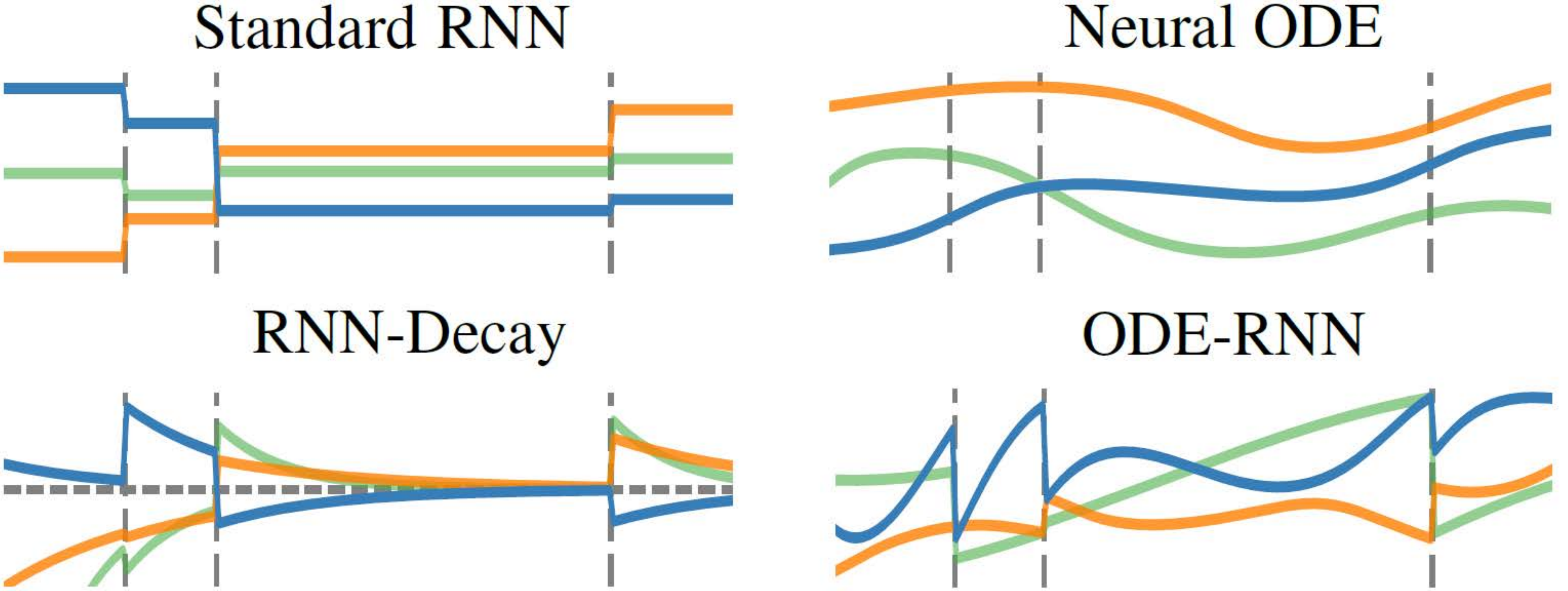} 
    \caption{Latent state trajectories. Vertical lines indicate sequential inputs; colored lines show dimensions of the latent state. Figure adapted from \cite{rubanova2019latent}.}
    \label{fig:ode}
\end{figure}

The recurrent neural network (RNN; originally introduced by \cite{rumelhart1986learning}) has been widely adopted to model sequential and temporal information. 
In a vanilla RNN structure, each RNN cell encodes sequential inputs iteratively into a latent state, where the previous output is used as input to the next iteration. 
This characteristic empowers the model to capture relations between consecutive inputs of a sequence and enables it to accumulate an overall understanding of the whole trajectory. 

However, while the vanilla RNN (and its variants with discrete updating of the latent state) have achieved state-of-the-art performance in tasks such as time series modeling \cite{connor1994recurrent} and natural language processing \cite{mikolov2010recurrent,sutskever2014sequence}, they do not perform well on irregularly sampled time series (i.e., where the time intervals between sequential inputs are varied).
To simulate the influence of time, one option is to model the latent state continuously between sequential inputs. 
To achieve this, the RNN-Decay model \cite{che2018recurrent,mei2017neural} decays the latent state between sequential inputs with a pre-defined exponential kernel, such that the latent state tends to be deactivated gradually. However, as the decay function is pre-defined, the RNN-Decay model risks under-fitting. 

The ODE-RNN \cite{rubanova2019latent,Habiba2020} is an extension of Neural ODE (originally proposed by \cite{chen2018neural}), characterized by a continuous latent state representation where the evolution process is learned rather than pre-defined. 
ODE-RNN models the first-order derivative of the latent state over observed steps instead of its exact state function, therefore generalizing discrete updates in neural networks to continuous dynamics. A comparison of latent state trajectories in RNN schemes is illustrated in Fig.~\ref{fig:ode}.

In ODE-RNN, a hidden state $h(t)$ is defined as a solution to an ODE initial value problem:

\begin{equation}
    \begin{array}{cl}
    \frac{dh(t)}{dt}=f_{\theta }(h(t),t) \; \mathrm{where} \; h(t_{0})=h_{0}
    \end{array}
    \label{eqt_1}
\end{equation}

where function $f_{\theta }$ is a neural network simulating latent state dynamics with parameter set $\theta$. 
The gradients can be derived by the adjoint sensitivity method \cite{pontryagin1962}. 
Specifically, in ODE-RNN, neural ODE modules are inserted between sequential inputs, with GRU gating used to control the instant transformation of latent state at exact observation time points. The latent state between two inputs can then be derived using an ODE solver as:

\begin{equation}
    \begin{array}{cl}
        {h_{i}}'=\mathrm{ODEsolver}(f_{\theta},h_{i-1},(t_{i-1},t_{i}))
    \end{array}
    \label{eqt_2}
\end{equation}

The superiority of the ODE-RNN model over traditional RNN models with discrete updates results from the ability to learn the evolution dynamics of the latent state for any given length of time intervals. 
ODE-RNN has demonstrated enhanced accuracy over discrete RNN schemes in datasets with irregularly sampled observations such as medical records and human movements  \cite{rubanova2019latent}. For this reason, we consider the ODE-RNN for the LOBRM.

\section{The LOB Recreation Model (LOBRM)}\label{sec:LOB-model}

\begin{figure*}[tb]
\centering
\includegraphics[width=0.95\linewidth]{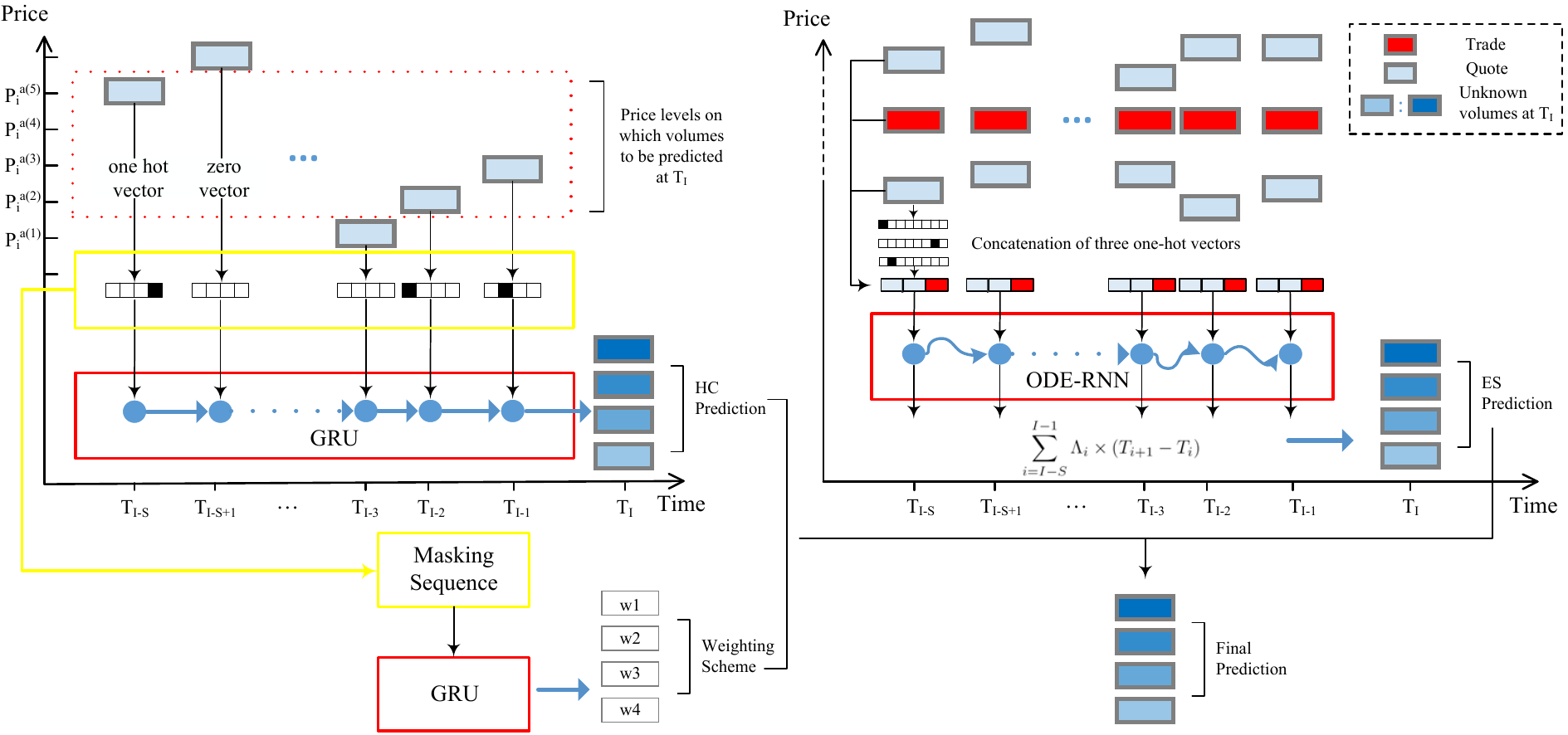} 
\caption{Overview of LOBRM for the ask side. The bid side prediction follows a similar workflow.}
\label{fig_model}
\end{figure*}

\subsection{Problem Description} 
We attempt to recreate the LOB with 5 price levels for small-tick stocks using TAQ data. 
Empirical studies on small-tick stocks suggest that orders in the LOB tend to be densely distributed around the top price levels, with one tick (the minimum price interval) between each level \cite{blanchet2017unraveling}. We confirm a similar distribution is observed in the LOBSTER data for MSFT and INTC.
Therefore, as the price for the current top levels of the LOB can be directly observed from TAQ data, the price for deeper levels can be deduced by simply adding or subtracting ticks. This simplification reduces the problem to predicting order volumes in the hidden levels of the LOB. 

For generalization, we denote trades and quotes streams as $\left \{ TD_{i} \right \}_{i\in \textsl{n}}$ and $\left \{ QT_{i} \right \}_{i\in \textsl{n}}$ respectively, and trajectories of time points for TAQ records as $\left \{ T_{i} \right \}_{i\in \textsl{n}}$, indexed by $\textsl{n}=\left \{ 1,\ldots,N \right \}$, where $N$ equals the number of time points in TAQ data. The states of the LOB sampled at $\left \{ T_{i} \right \}_{i\in \textsl{n}}$ are denoted as $\left \{ LOB_{i} \right \}_{i\in \textsl{n}}$. For each record at time $T_{i}$, $QT_{i}=( p_{i}^{a(1)},v_{i}^{a(1)},p_{i}^{b(1)},v_{i}^{b(1)})$, where $p_{i}^{a(1)},v_{i}^{a(1)},p_{i}^{b(1)},v_{i}^{b(1)}$ denote best ask price, order volume at best ask, best bid price, and order volume at best bid, respectively. $TD_{i}=(p_{i}^{td},v_{i}^{td}, d_{i}^{td})$, where $p_{i}^{td},v_{i}^{td},d_{i}^{td}$ denote price, volume, and direction of the trade, with $+1$ and $-1$ indicating orders being sold or bought. $LOB_{i}=(p_{i}^{a(l)},v_{i}^{a(l)},p_{i}^{b(l)},v_{i}^{b(l)})$ depicts the price and volume information at all price levels, with $l\in(1,...,L)$. 

As we focus on a LOB with 5 price levels, hence $L=5$. $QT_{i}$ denotes the top levels of $LOB_{i}$. From the aforementioned empirical results, we have $p_{i}^{a(l)}=p_{i}^{a(1)}+(l-1)*tick size$ and $p_{i}^{b(l)}=p_{i}^{b(1)}-(l-1)*tick size$. Under this formulation, for a single sample, the model predicts $(v_{I}^{a(2)},...,v_{I}^{a(L)})$ and $(v_{I}^{b(2)},...,v_{I}^{b(L)})$ conditioned on the observations of $\left \{ QT_{i} \right \}_{{I-S}:I}$ and $\left \{ TD_{i} \right \}_{{I-S}:I}$, with $S$ being the number of time steps that the model looks back in TAQ data history. 

\subsection{Model Structure}
An overview of the model structure is presented in Fig.~\ref{fig_model}. As a common practice in exploiting LOB, the ask side and bid side of the LOB are modelled separately. Here we only illustrate the modelling of the ask side, as the modelling of the bid side follows exactly the same logic. We first illustrate the encoding method used for TAQ data and then describe the three components of the LOBRM: the history compiler, the market events simulator, and the adaptive weighting scheme. 

\subsection{One-Hot Positional Encoding for TAQ data}
We use a sparse one-hot vector encoding to extract features from TAQ records, with volume encoded explicitly as an element in the feature vector and price level encoded implicitly by the position of the element. 

In the history compiler, we consider only past volume information at current deep price levels. Conditioned on the best ask price $p_{I}^{a(1)}$ at target time $T_{I}$, let $sp_{s}^{a}=(p_{I-s}^{a(1)}-p_{I}^{a(1)})/ticksize$ be the distance between a history ask quote and current best ask price. 

We represent an ask quote record $(p_{I-s}^{a(1)},v_{I-s}^{a(1)})$ as:
\begin{equation}
    \left\{
    \begin{array}{cl}
    O_{L-1}, \mathrm{where}\,o_{sp_{s}^{a}}=v_{I-s}^{a(1)} & \mathrm{if} \,sp_{s}^{a}\in[1,L-1] \\ Z_{L-1} & \mathrm{otherwise}
    \end{array} \right.
    \label{eqt_3}
\end{equation}
where $O_{L-1}$ is a one-hot vector with dimension $1\times (L-1)$; $o_{sp_{s}^{a}}$ denotes the  
${sp_{s}^{a}}$-th element of the vector; and  $Z_{L-1}$ denotes a zero vector of the same dimension. Then, the corresponding mask used later in the weighting scheme is denoted as:
\begin{equation}
    \left\{
    \begin{array}{cl}
    O_{L-1}, \mathrm{where}\,o_{sp_{s}^{a}}=1 & \mathrm{if} \,sp_{s}^{a}\in[1,L-1] \\ Z_{L-1} &  \mathrm{otherwise}
    \end{array} \right.
    \label{eqt_4}
\end{equation}

In the market events simulator, the model perceives the whole trajectory of TAQ data at all price levels. Conditioned on current best ask price $p_{I}^{a(1)}$ at $T_{I}$, we represent an ask quote record $(p_{I-s}^{a(1)},v_{I-s}^{a(1)})$ at $T_{I-s}$ as:
\begin{equation}
    \begin{array}{cl}
        O_{2k-1}^{aq}, \mathrm{where}\,o_{k+sp_{s}^{a}}=v_{I-s}^{a(1)}
    \end{array}
    \label{eqt_5}
\end{equation}
where $k\in R$ and $2k-1>>L-1$. 
For a bid quote record $(p_{I-s}^{b(1)},v_{I-s}^{b(1)})$, we have $sp_{s}^{b}=(p_{I-s}^{b(1)}-p_{I}^{b(1)})/ticksize$. Thus, we represent a bid quote record as:
\begin{equation}
    \begin{array}{cl}
        O_{2k-1}^{bq}, \mathrm{where}\,o_{k+sp_{s}^{b}}=v_{I-s}^{b(1)}
    \end{array}
    \label{eqt_6}
\end{equation}
and a trade record $(p_{I-s}^{td},v_{I-s}^{td}, d_{I-s}^{td})$, is represented as:
\begin{equation}
    \left\{
    \begin{array}{cl}
        O_{2k-1}^{td}, \mathrm{where}\,o_{p_{I-s}^{td}-p_{I}^{a(1)}}=v_{I-s}^{td} & \mathrm{if} \,d_{I-s}^{td}<0 \\
        O_{2k-1}^{td}, \mathrm{where}\,o_{p_{I-s}^{td}-p_{I}^{b(1)}}=v_{I-s}^{td} & \mathrm{if} \,d_{I-s}^{td}>0   \\
    \end{array} \right.
    \label{eqt_7}
\end{equation}

Finally, those three features are concatenated and are used as input into the market events simulator. We find in experiments that the one-hot positional encoding method for TAQ data is robust, as it remains sparse while including all information in the TAQ trajectory.

\subsection{The History Compiler (HC)}
The HC predicts the LOB from a historical perspective. 
It contains a GRU module to compile volume histories from the ask quotes trajectory, only if quote price is among $(p_{I}^{a(2)},...,p_{I}^{a(L)})$ at time $T_{I}$.
As quote price fluctuates over time, it is likely that historical quote prices overlap with the top price levels at target time, therefore the volume information of these historical records tells us how many orders were previously sitting at those price levels. 
In this sense, quote histories can be used for deducing a rough estimation of the current LOB status, despite the fact that orders are frequently submitted and cancelled.

\subsection{The Market Events Simulator (ES)}

The ES predicts the LOB from a dynamic perspective. As market participants, particularly algorithmic trading systems, submit and cancel orders at millisecond granularity, we model arrival of limit orders at different price levels as a probabilistic process by the means of continuous RNNs. Following prior probabilistic modelling of the LOB, we assume the occurrences of market events follow independent Poisson processes with time-varying arrival rates, and consequently their aggregation effect on net order arrivals still follows a pooled inhomegeneous Poisson process.

We derive the vector of net order arrival rates $\Lambda_{I-s}=[\lambda_{I-s}^{a(2)},...,\lambda_{I-s}^{a(L)} ]$ for different price levels at time $T_{I-s}$ directly from the latent state $h_{I-s}$ using MLP layers as:
\begin{equation}
    h_{I-s}=\mathrm{ODEsolver}(f_{\theta},h_{I-s-1},(t_{I-s-1},t_{I-s}))
\end{equation}
\begin{equation}
    \Lambda_{I-s}=\mathrm{MLP}(h_{I-s})
\end{equation}
where $f_{\theta}$ is a neural network parameterized by $\theta$, which is learned in a data-driven manner to deduce the derivative of the latent state. After acquiring the trajectory of $\Lambda$ at all trade times over the defined length of time steps, we calculate the accumulated order volumes between $[T_{I-S},T_{I}]$ as:
\begin{equation}
    \sum_{i=I-S}^{I-1}\Lambda_i\times (T_{i+1}-T_{i})
\end{equation}

\subsection{The Weighting Scheme (WS)}
We use an adaptive weighting scheme to combine the HC and ES predictions. 
A weighting vector with length equal to the number of price levels on which volumes are predicted (here equal to four) is derived from the masking sequence of ask quotes used in the HC. 
The masking sequence of ask quotes history indicates the timing and abundancy of records at each price level, and a higher weighting is given to price levels with a recent and abundant quote history as this information is more reliable. 
We again use a GRU module to encode the masking sequence, and the decoded information after sigmoid activation indicates the reliability of the HC prediction at each price level.  
By balancing the predictions according to the weighting scheme, the model produces its final prediction of order volumes at $p_{I}^{a(2)},...,p_{I}^{a(L)}$.

\section{Empirical Analysis of LOBRM}\label{sec:experiments}

We present an empirical analysis of LOBRM on real world LOB data provided by LOBSTER for one full day of trading (12/06/2012) for two small-tick technology stocks (Microsoft, symbol \textit{MSFT}; and Intel, symbol \textit{INTC}).\footnote{\url{https://lobsterdata.com/info/DataSamples.php}}
These data contain $\approx$1 million LOB updates. 

\subsection{Data Preprocessing}
\begin{figure}[tb]
\centering
\includegraphics[width=0.8\linewidth]{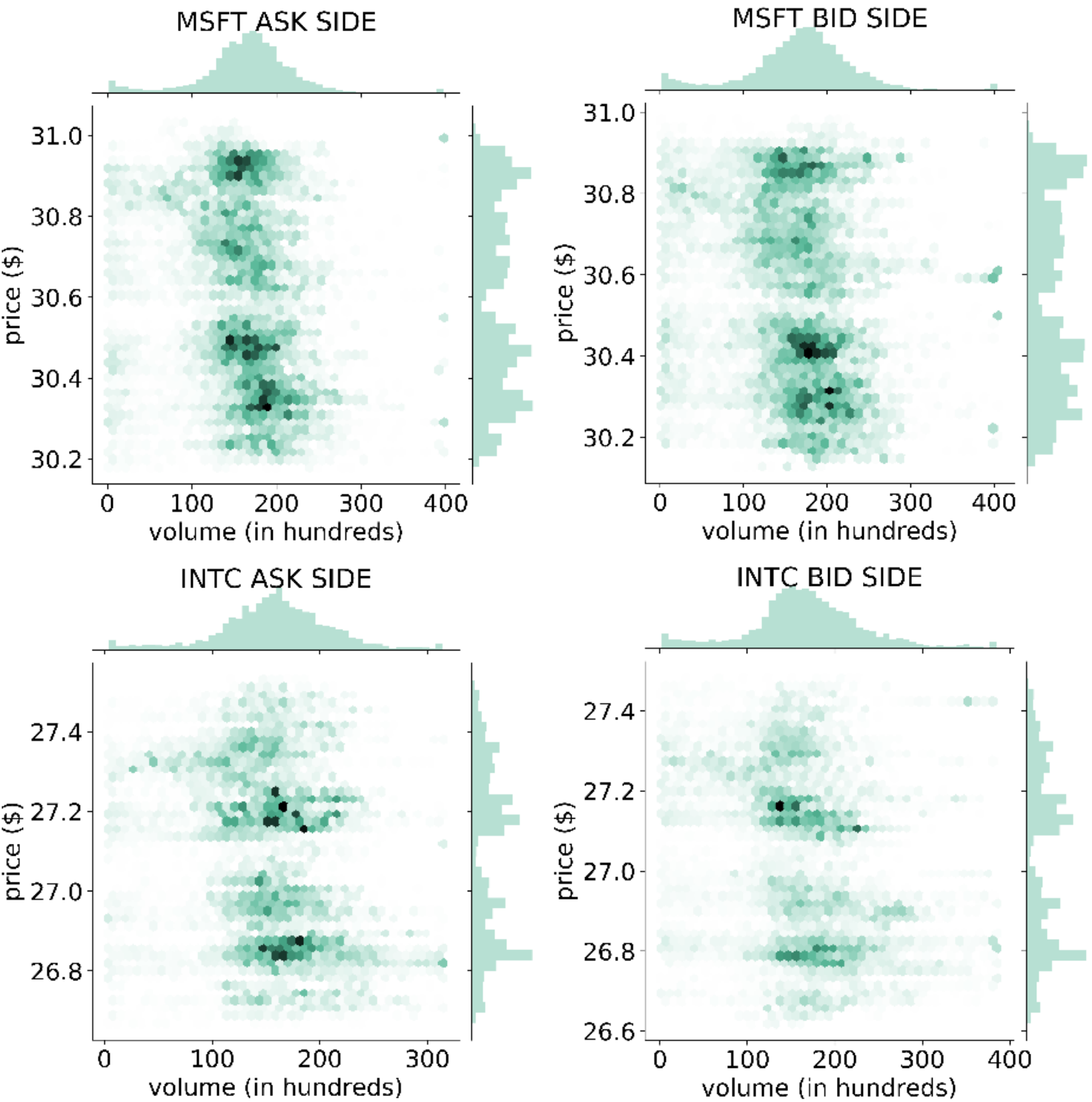} 
\caption{2-D distribution of the dataset after winsorization.}
\label{fig:dist}
\end{figure}

Following \cite{blanchet2017unraveling}, we select unique time points on which trades happen for LOB recreation.
As a common practice to reduce noise, we also remove LOB data during the first half-hour after market open and the last half-hour before market close. 
This process results in 10K time series samples over 5.5 hours of trading.  
We then directly extract TAQ data from the LOB and its affiliated message book. 
To alleviate the effect of outliers, we divide all volume numbers by 100 and winsorize the data by the range $[0.005,0.995]$. The processed data distribution is shown in  Fig.~\ref{fig:dist}. We set parameters $S=100$ and $k=8$. After testing $S=\{50,100,150\}$, we found $S=100$ has best performance; while $k$ is selected to cover more than 90\% of observed price movements in window of size $S$.

For training and testing the source model, we extract TAQ (top line) and LOB data from the full LOBSTER dataset. These data are then converted to time series samples using a rolling window of size $S=100$, such that the first sample consists of TAQ histories at timesteps 1-100 and is labeled by LOB volume at deep price levels at time step 100. The second sample consists of TAQ history at timesteps 2-101 and is labeled using LOB volume at deep price levels at time step 101, etc. Samples are then shuffled (producing a random ordering) and split into train (80\%) and test (20\%) sets.

\begin{table*}[tb]
\caption{Comparison between models. Introducing non-linearity and temporal continuity significantly enhance LOB prediction.} 
\footnotesize
\centering
\begin{tabular}{@{}lcSSSSSSSc@{}}
\toprule
        \multirow{2}{*}{Model} & \multirow{2}{*}{Type} & \multicolumn{2}{c}{Train Loss}  
        & \multicolumn{2}{c}{Test Loss}
        & \multicolumn{2}{c}{Test Loss (\%)} 
        & \multicolumn{2}{c}{$R^2$}\\ 

        \rule{0pt}{2ex}  
        && \multicolumn{1}{c}{Bid} & \multicolumn{1}{c}{Ask} 
        & \multicolumn{1}{c}{Bid} & \multicolumn{1}{c}{Ask} 
        & \multicolumn{1}{c}{Bid} & \multicolumn{1}{c}{Ask} 
        & \multicolumn{1}{c}{Bid} & \multicolumn{1}{c}{Ask}\\
\midrule
RR     & (Linear) &  30.94   &  26.94    &  32.03  &  27.04  &   17.32\%    &   15.22\%    &  0.169 &  0.141 \\ 
SVR     & (Linear) &  30.93   &  26.92    &  31.99  &  26.97  &   17.30\%    &   15.18\%    &  0.167 &  0.145 \\ 
RF     & (Non-Linear) &  7.29   &  6.23    &  19.45  &  15.98  &   10.52\%    &   8.99\%    &  0.638 &  0.641 \\ 
SLFN     & (Non-Linear) &  5.88   &  5.40    &  18.62  &  16.43  &   10.07\%    &   9.25\%    &  0.627 &  0.609 \\ 
LOBRM (GRU)     & (Discrete RNN) &  10.95   &  9.69    &  15.44  &  13.56  &   8.35\%    &   7.63\%    &  0.726 &  0.675 \\ 
LOBRM (GRU-T)   & (Discrete RNN) &  11.96   &  9.28    &  16.96  &  13.17  &   9.17\%    &   7.41\%    &  0.687 &  0.688 \\ 
LOBRM (LSTM)    & (Discrete RNN) &  11.91   &  9.38    &  17.28  &  13.83  &   9.35\%    &   7.79\%    &  0.674 &  0.681 \\ 
LOBRM (LSTM-T)  & (Discrete RNN) &  11.77   &  9.93    &  15.99  &  13.86  &   8.65\%    &   7.80\%    &  0.695 &  0.675 \\ 
LOBRM (ODE-RNN) & (Continuous RNN) &  7.22    &  6.62    &  13.61  &  11.56  &   7.36\%    &   6.50\%    &  0.773 &  0.753 \\ 
\bottomrule
\end{tabular}
\label{tab:continuous}
\end{table*}

\begin{table*}[tb]
\caption{Ablation study (HC = History Compiler; ES = Event Simulator; WS = Weighting Scheme).}
\footnotesize
\centering
\begin{tabular}{@{}lSSSSSSSc@{}}
\toprule
         & \multicolumn{2}{c}{Train Loss} 
         & \multicolumn{2}{c}{Test Loss} 
         & \multicolumn{2}{c}{Test Loss (\%)} 
         & \multicolumn{2}{c}{$R^2$}\\ 
         
         \rule{0pt}{2ex}
         & {Bid} & {Ask} & {Bid} & {Ask} & {Bid} & {Ask} & {Bid} & {Ask} \\
\midrule
HC       &  26.44   &  22.73   &  28.10  &  23.33  &   15.20\%   &   12.57\%   &  0.273 &  0.291 \\
ES       &  7.51    &  6.91    &  13.83  &  12.44  &   7.48\%    &   7.00\%    &  0.769 &  0.723 \\
HC + ES  &  7.71    &  7.06    &  13.36  &  11.62  &   7.23 \%   &   6.54\%    &  0.775 &  0.743 \\
HC + ES + WS &  7.22    &  6.62    &  13.61  &  11.56  &   7.36\%    &   6.50\%    &  0.773 &  0.753 \\
\bottomrule
\end{tabular}
\label{tab:ablation}
\end{table*}

\subsection{I. Comparison of Models in LOB prediction}\label{sec:exp1}

 We first illustrate LOB prediction performance of mainstream regression and machine learning methods: (1) Ridge Regression (RR); (2) Support Vector Regression (SVR); (3) Random Forest (RF); and (4) Single Layer Feedforward Network (SLFN). We then evaluate the performance of LOBRMs with ODE-RNN substituted by four discrete RNN modules: (1) LSTM; (2) LSTM-T, with time concatenated input; (3) GRU; and (4) GRU-T, with time concatenated input. Finally, we show the performance of the full LOBRM.

In the HC and ES, latent size is set to 32. Decoders consist of two MLP layers containing 64 units, with LeakyReLU and Tanh activation respectively. In the WS, latent size is set to 16, and the decoder consists of one MLP layer containing 16 units with Sigmoid activation. The network used to derive the derivative of the latent state consists of three MLP layers containing 64 units with Tanh activation. The Euler method is used to solve differential equations for ODE-RNN. Batch size is set to 64 and loss function is L1 loss. We use 80\% of the shuffled data to train the model. The initial learning rate is set to 1e-2 and it gradually decays to 1e-3, with a decay rate 0.999. It roughly takes 1000 and 250 iterations for discrete and continuous LOBRMs to converge.

To evaluate these models, three criteria are used: (1) L1 loss function is used to measure the average absolute distance between the predicted volume and the ground truth volume; (2) L1 loss/average volume in ground truth, which measures the prediction accuracy as a percentage; and (3) R-squared, calculated on the test data, using the method presented by \cite{blanchet2017unraveling} to gain a better comparison with the existing literature. 

Table~\ref{tab:continuous} presents evaluation results, and demonstrates that the LOBRM with ODE-RNN module outperforms alternative discrete RNN modules on all criteria. 
Loss curves (not shown) indicate that the LOBRM with continuous ODE-RNN is more efficient, reaching a lower loss value after 50 iterations than the discrete RNNs reach after 1000 iterations.
We find that the concatenation of time onto feature vectors in discrete RNN models has little effect on the model performance. Additionally, we observe that non-linear models overwhelmingly outperform linear models, the fact of which is consistent with \cite{sirignano2019universal}.

To the authors' knowledge, this is the first deep learning attempt to address the problem of recreating the LOB, therefore we are unable to compare results with the deep learning literature. 
While statistical models have previously been used to approach a similar problem, results comparison is not ideal as either the evaluation criteria are different or models are for different time scales.
The most similar study is by \cite{blanchet2017unraveling}, who present R-squared values in the range $[0.81,0.88]$ for daily average volume at different price levels. However, in comparison, our R-squared values presented in Table~\ref{tab:continuous} are regressed against the ground truth with no averaging procedure. When we average the predictions over every batch at each price level, which transforms volume predictions into approximately five minutes frequency, the value of R-squared is over 0.9. This value would rise again if our predictions are averaged into daily frequency. Therefore, we suggest performance of the full LOBRM is superior than related work in the literature. 

\subsection{II. Ablation Study}\label{sec:exp2}

We conduct an ablation study to investigate the contribution to prediction accuracy of each module in the LOBRM. Four experiments are conducted: HC, using only the history compiler; ES, using only the market events simulator; HC+ES, which includes the history compiler and event simulator, using a pre-defined weighting scheme to combine outputs; and HC+ES+WS, which is the full LOBRM with adaptive weighting scheme. Model specifications are the same as Experiment~I, except all model training lasts 250 iterations.

Results from the ablation study are presented in Table~\ref{tab:ablation}.
We see that predictions from the HC alone have relatively high error. In comparison, the ES achieves much higher performance, and demonstrates that the market events simulator, dominated by the ODE-RNN, is the most important component in the LOBRM. Finally, combining the predictions using either the pre-defined or adaptive weighting scheme achieves the best test performance, suggesting that the history compiler can add benefit when used in conjunction with the event simulator. We conduct Wilcoxon signed rank tests on the test L1 loss between different experiment sets to test whether the improvement is statistically significant. We find that ES has lower test loss than HC (p-value\textless0.01, bid/ask side) and HC+ES has lower test loss than ES (p-value\textless0.01, bid/ask side). Although the WS module contributes little to accuracy, it makes the full LOBRM more flexible and we find in later experiments that its transferred model performs better. Therefore, we proceed with the full LOBRM as our main model.

\subsection{III. Transfer Learning}\label{sec:exp3}

We attempt transfer learning \cite{pratt1993discriminability,pan2009survey} by using the full LOBRM trained on one stock to perform prediction on a different stock. We first learn the source model on the MSFT dataset and then fine tune the model for INTC, using only 30\% of INTC data for the fine tuning. The model specifications used are the same as Experiment~I, except that batch size, final learning rate, and number of iterations are set as 32, 5e-4, and 500 respectively.

\begin{table}[tb]
\caption{Transfer learning results.}
\small
\centering
\begin{tabular}{@{}lSSSc@{}}
\toprule
         & {Train Loss} & {Test Loss} & {Test Loss (\%)} & {$R^2$}\\ 
\midrule
Bid Side &  3.96    &  17.70  &   9.78\%    &  0.685 \\ 
Ask Side &  4.30    &  15.04  &   8.95\%    &  0.649 \\ 
\bottomrule
\end{tabular}
\label{tab:transfer}
\end{table}

\begin{figure}[tb]
  \centering
  \includegraphics[width=0.72\linewidth]{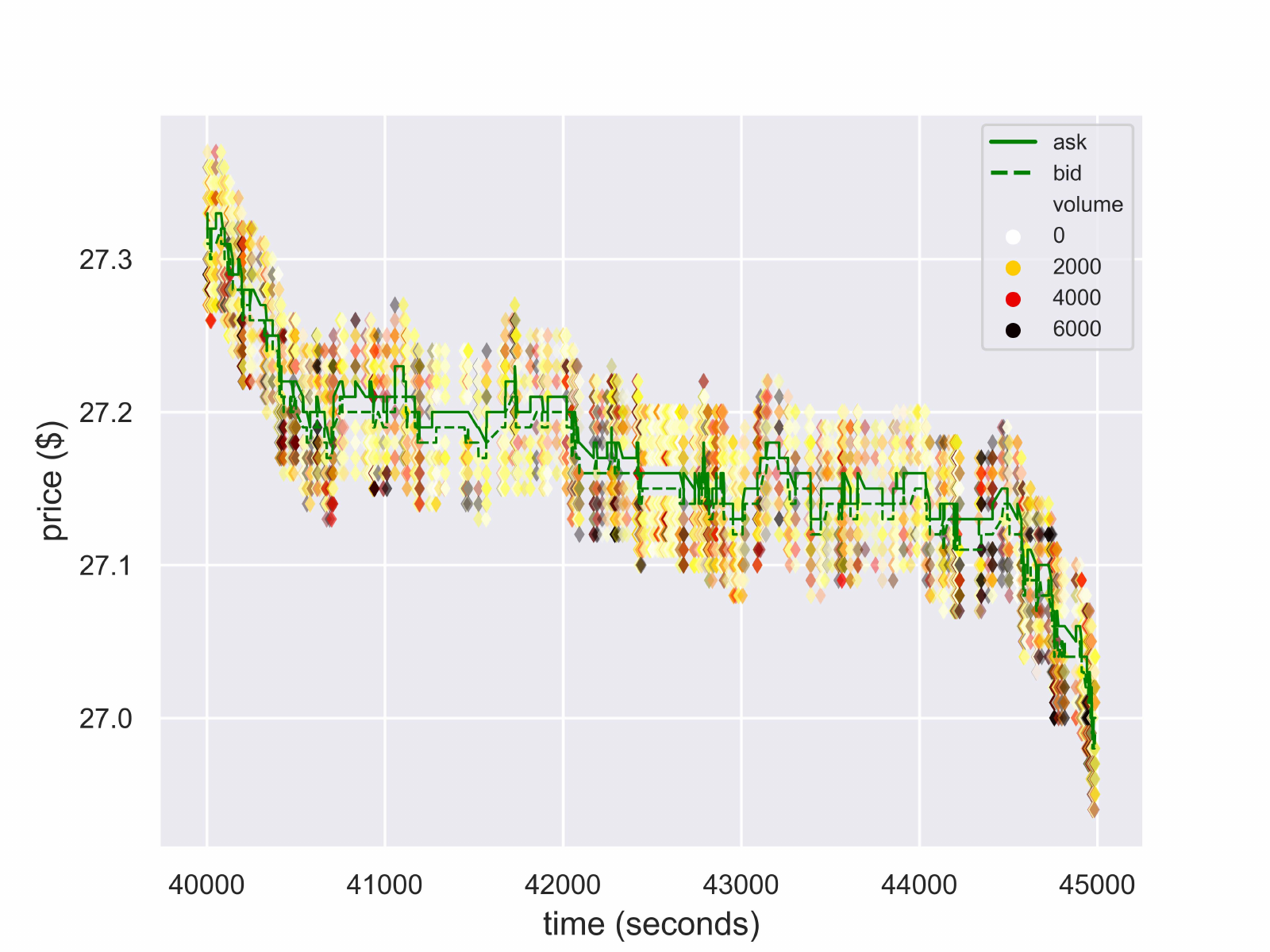}
  \caption{Visualization of the difference between order volumes in the real and the recreated LOB for symbol INTC. Colors indicate volume differences; time is seconds since midnight. Average volume across all price levels $\approx17,500$.}
  \label{fig:LOB-heatmap}
\end{figure}

Results are shown in Table~\ref{tab:transfer}. We see that the accuracy of the transferred model for INTC (Table~\ref{tab:transfer}) is lower than the accuracy of the fully trained model for MSFT (Table~\ref{tab:continuous}). However, the transferred model reaches an accuracy that is approximately equivalent to a fully trained model using discrete RNNs (Table~\ref{tab:continuous}).

Fig.~\ref{fig:LOB-heatmap} presents a visualisation of the difference between the real and the recreated LOB for INTC, during a random time period.
We see that the differences are mostly under 2000 units, which demonstrates the utility of the transferred model and provides evidence that it is possible to apply the LOBRM to financial assets with limited LOB training data. 

\subsection{IV: Application Scenario}

\begin{table}[tb]
\caption{Mid-price movement prediction results.}
\footnotesize
\centering
\begin{tabular}{@{}lSS@{}}
\toprule
{LOB Data} & {Train Accuracy} & {Test Accuracy} \\ 
\midrule
Top level only (Quote) &  84.05\%      &  75.17\%     \\
Real LOB (LOBSTER) &  93.94\%      &  81.14\%     \\
Recreated LOB (LOBRM) & 93.93\%      &  79.75\%     \\ 
\bottomrule
\end{tabular}
\label{tab:mid-price}
\end{table}

Finally, we provide a potential application scenario of the recreated LOB for INTC in order to demonstrate that the recreated LOB contains the majority of information in the real LOB. 
For market practitioners, LOB data is a valuable source of information for predicting future mid price trends. Here, we compare the future mid-price prediction accuracy of using: (i) the recreated LOB; (ii) the real LOB; and (iii) only the first level of LOB (equivalent to using quote data only). The model attempts to predict the mid-price movement at the next time point as \textit{up}, \textit{same}, or \textit{down}. To predict mid-price, we follow the method proposed by \cite{zhang2019deeplob}.

The mid price prediction model has two convolution layers with 16 $[1*2]$ kernels and $[1*2]$ strides in the first layer, and 16 $[1*5]$ kernels and $[1*5]$ strides in the second layer. 
These settings are chosen to fit the structure of the LOB. Both layers are followed by batch normalization layers and LeakyReLU activation. The outputs from the convolutional layers are used as sequential inputs into a GRU module. To enhance accuracy, we replace the inception module with a temporal attention module \cite{luong2015effective}. The output is activated by a Softmax function to produce a probabilistic distribution over three trend labels. For data preprocessing, we run a rolling average of five time steps to alleviate label imbalance \cite{ntakaris2018benchmark}. Label distribution after preprocessing is approximately balanced, with 31\%, 39\%, and 30\% for \textit{up}, \textit{same}, and \textit{down}, respectively. We train the model with cross entropy loss and run experiments for 200 iterations. 

Table~\ref{tab:mid-price} presents results. 
We see that using only the top level of the LOB achieves much lower accuracy than using a real LOB with five price levels. 
This result conforms with the finding that the top level accounts for $\approx80\%$ of future price movements \cite{cao2009information}. 
The real LOB accuracy of 81.14\% is also consistent with the accuracy range $[0.75,0.84]$ for the same three-class classification problem presented by \cite{zhang2019deeplob}, and higher than the range $[0.65, 0.76]$ for the two-class problem (\textit{up}, \textit{down}) presented by \cite{sirignano2019universal}.
More pertinently, the LOBRM has only 1.5\% lower accuracy than the real LOB, and 4.5\% higher accuracy than the top level, demonstrating that the LOBRM can recover the majority of LOB information.

\section{Conclusion}\label{conclusions}

We have presented the LOB recreation model (LOBRM) to recover the LOB of five price levels for small-tick stocks from only TAQ data. To the authors' knowledge, this is the first attempt to solve this problem from a deep learning perspective. 
The LOBRM contains three components: a history compiler, a market events simulator, and an adaptive weighting scheme. We have demonstrated accuracy is improved by encoding irregularly sampled TAQ data into a continuous latent state using an ODE-RNN, compared with using discrete RNN variants. Through an ablation study, we find that even though the market events simulator plays a dominant role in LOB prediction accuracy, the combination of all three components further improves accuracy. 
The results of applying transfer learning show that the knowledge learned on LOB data for one stock can be transferred to a different stock with a relatively small amount of additional data needed for fine tuning. Finally, we demonstrated that the recreated LOB can be effectively applied to the real world scenario of predicting future mid-price movements.
The LOBRM enables us to create a synthetic LOB at no extra cost from only TAQ data. In cases where either the historical LOB records are incomplete, or online streaming of the LOB data is prohibitively expensive, the LOBRM can be a valuable tool for practitioners and researchers alike.

\paragraph{Limitations and Future Work}
Since we also suffer from the scarce availability of public LOB data, an obvious limitation of this research is that we only train and test the LOBRM on two intraday datasets. 
There exist LOB datasets with longer time horizons, such as the FI-2010 dataset \cite{ntakaris2018benchmark}, but unfortunately this dataset is missing timestamps. 
We have recently sourced additional LOBSTER data and will continue to conduct larger scale research. We also plan to use the finding of 
\cite{sirignano2019universal} to attempt a universal LOBRM model; and will incorporate the use of simulation platforms to model LOB recreation and mid-price prediction for automated trading. 

\section*{Acknowledgements}\label{ack}
Zijian Shi's PhD is supported by a China Scholarship Council (CSC)/University of Bristol joint-funded scholarship. John Cartlidge is sponsored by Refinitiv.

\ifnum\LNCSmode=0
  \bibliographystyle{ACM-Reference-Format}
\else
  \bibliographystyle{splncs03}
\fi


\end{document}